# Engineering spin-orbital magnetic insulator by tailoring superlattices

Jobu Matsuno[1*], Kota Ihara[2], Shugen Yamamura[3], Hiroki Wadati[3], Kenji Ishii[4], V. Vijay Shankar[5], Hae-Young Kee[5,6], and Hidenori Takagi[1,7]

[1]*RIKEN Advanced Science Institute, Wako, Saitama 351-0198, Japan*

[2]*Department of Advanced Materials Science, University of Tokyo, Kashiwa, Chiba 277-8561, Japan*

[3]*Department of Applied Physics and Quantum-Phase Electronics Center (QPEC), University of Tokyo, Hongo, Tokyo 113-8656, Japan*

[4]*SPring-8, Japan Atomic Energy Agency, Sayo, Hyogo 679-5148, Japan*

[5]*Department of Physics, University of Toronto, Toronto, Ontario M5S 1A7, Canada*

[6]*Canadian Institute for Advanced Research, Toronto, Ontario M5G 1Z8, Canada*

[7]*Department of Physics, University of Tokyo, Hongo, Tokyo 113-0033, Japan*

RIKEN Advanced Science Institute, 2-1 Hirosawa, Wako, Saitama 351-0198, Japan

TEL: +81-48-467-9348, FAX: +81-48-462-4649

Correspondence should be addressed to J.M. (matsuno@riken.jp) and H.T. (takagi@qmat.phys.s.u-tokyo.ac.jp).

[*]Present address: RIKEN Center for Emergent Matter Science (CEMS), Wako, Saitama 351-0198, Japan

[*]Present address: RIKEN Center for Emergent Matter Science (CEMS), Wako, Saitama 351-0198, Japan

**Novel interplay of spin-orbit coupling and electron correlations in complex Ir oxides recently emerged as a new paradigm for correlated electron physics. Because of a large spin-orbit coupling of ~0.5 eV, which is comparable to the transfer energy $t$ and the crystal field splitting $\Delta$ and Coulomb $U$, a variety of ground states including magnetic insulator, band insulator, semimetal and metal, shows up in a narrow materials phase space. Utilizing such subtle competition of the ground states, we successfully tailor a spin-orbital magnetic insulator out of a semimetal $SrIrO_3$ by controlling dimensionality using superlattice of $[(SrIrO_3)_m, SrTiO_3]$ and show that a magnetic ordering triggers the transition to magnetic insulator. Those results can be described well by a first-principles calculation. This study is an important step towards the design and the realization of topological phases in complex Ir oxides with very strong spin-orbit coupling.**

One of the appealing examples of exotic electronic phases produced by the novel interplay of spin-orbit coupling (SOC) and the other relevant parameters in complex Ir oxides is a novel spin-orbital Mott insulator associated with the half-filled $J_{eff} = 1/2$ band, lately identified in layered perovskite $Sr_2IrO_4$[1,2]. The three dimensional counterpart of $Sr_2IrO_4$, $SrIrO_3$, is revealed to be a semimetal close to a band insulator[3,4], where a line of Dirac nodes generated by combination of SOC and lattice symmetry plays a key role[5]. The large SOC may give rise to intriguing topological phases as well in their sister iridium oxides. It was theoretically predicted that iridium oxides with unique geometry of lattice, such as honeycomb or pyrochlore, may have a topological character[6–8]. A family of complex iridium oxides is thus a promising playground to explore rich electronic phases, ranging from semimetal, magnetic insulator and even topological insulator, by the subtle control of lattice.

The control of dimensionality and the introduction of interface using superlattice structure have been demonstrated to be a useful technique to control the electronic phase of $3d$ transition metal oxides including titanium[9] and nickel[10]. The approach using superlattice could be even more powerful in complex $5d$ Ir oxides than their $3d$ analogues to explore exotic electronic phases, because of the presence of the large SOC. The emergence of a variety of phases in a narrow materials space, as a consequence of the interplay between SOC and other electronic energy scales, means that only a minute change of the control parameters including dimensionality may totally alter the ground states. To realize topological phases, the modification of local symmetry of lattice, for example breaking inversion symmetry, is often essential, which could be done by introducing interfaces. Indeed, a design of topological insulator was theoretically proposed by utilizing superlattice structure of perovskite oxides[5,11].

In this work, we have successfully grown artificial superlattices [$(SrIrO_3)_m$, $SrTiO_3$]/$SrTiO_3$(001) by pulsed laser deposition and made atomically thin slices of $SrIrO_3$ by inserting insulating $SrTiO_3$ ($d^0$). The dimensionality of the system by changing $m$ could be controlled systematically, which allowed us to explore the rich electronic phase diagram of the system.

The structural analysis with x-ray diffraction (XRD) measurements clarifies that the superlattices were fabricated as designed. Figure 1(a) shows the XRD scans (θ-2θ) of the superlattices with $m$ = 1, 2, 3, 4, and ∞. Besides the fundamental peaks originating from the cubic perovskite lattice ($a$ ~ 0.39 nm), clear superlattice peaks corresponding to the periodicity of $(m + 1) a$ were observed for all the superlattices. The in-plane lattice constants of superlattices matched well with that of the substrate as shown in Fig. 1(b), evidencing the coherent epitaxial growth. An atomically resolved HAADF (high-angle-angular-dark-field)-STEM (scanning transmission electron microscopy) image with strong atomic number ($Z$) contrast, is shown in Fig. 1(c) along the [100] direction of the SrTiO$_3$ substrate. Ir atoms indicated by the brightest spots are atomically aligned in the [001] plane. These results clearly demonstrate that the atomic order between Ir and Ti is indeed materialized in the coherently grown superlattices. We may view the superlattices as a tailored counterpart of Ruddlesden-Popper series, Sr$_{m+1}$Ir$_m$O$_{3m+1}$, consisting of stacked $m$-IrO$_2$ layers[12], which cannot be stabilized at ambient pressure when $m > 2$.

Reducing the dimensionality by the insertion of SrTiO$_3$ monolayer [see Fig. 2(a)], a (semi-)metal SrIrO$_3$ perovskite turned into an insulator. For pure SrIrO$_3$ film ($m = ∞$), a metallic behavior of resistivity $\rho(T)$, though poorly, was observed as shown in Fig. 2(b). Large and strongly temperature-dependent Hall effect in Fig. 2(c) is consistent with the semimetallic ground state of SrIrO$_3$, which was discussed to be the consequence of the presence of Dirac-like band[5]. With inserting SrTiO$_3$, ρ(*T*) systematically changes from metallic to insulating with decreasing the number of SrIrO$_3$ layers. *m* = 4 sample appears to be still (semi-)metallic. *m* = 3 is right at the border between metal and insulator. *m* = 2 and 1 appear to be an insulator. Closely inspecting the temperature dependence of ρ(*T*), we can see clear anomaly at around 110 K and 140 K for *m* = 2 and 1 film respectively. In the Hall effect, we do see a rapid increase on cooling below the temperature of anomaly, suggestive of the presence of transition to a strongly insulating state.

The magnetization measurement of those films clearly indicates that the (semi-)metal-insulator transition is triggered by the magnetic ordering. Fig. 2(d) establishes the link between the transport and the magnetism, where the in-plane magnetization *M*(*T*) measured at 0.1 T is demonstrated. We observe an in-plane weak ferromagnetism at low temperatures for the films with *m* ≤ 3 where an insulating ground state is realized. The onset of weak ferromagnetism coincides with the temperature where the transport anomaly associated with the transition to strong insulator is observed. Those close correlations between ρ(*T*) and *M*(*T*) indicate that the magnetism, rather than disorder, is a key ingredient for the (semi-)metal to insulator transition. We also note here that for *m* = 1 and 2, more significantly for *m* = 1, ρ(*T*) shows weak insulating behavior even above magnetic ordering, which might suggest the increased Mott character in the limit of reduced dimensionality.

The tailored magnetism in the superlattice turns out to be a canted in-plane antiferromagnetism. The weak ferromagnetic moments were observed only for magnetic field parallel to IrO$_2$ plane as indicated in Fig. 3(a). The weak in-plane moment of 0.075 $\mu_B$ was observed in the single layer Sr$_2$IrO$_4$, where the moment is produced by a Dzyaloshinskii-Moriya (DM) interaction associated with the rotation of IrO$_6$ octahedra[2] [see Fig. 4(c)]. It might be natural to consider that a similar canted in-plane antiferromagnetism is realized in the thin films. The resonant magnetic x-ray diffraction at Ir $L_3$ edge was measured for the $m = 1$ sample with σ-π' polarization to extract magnetic contribution. The (0.5, 0.5, 5) peak was clearly observed as shown in Fig. 3(b), consistent with the in-plane antiferromagnetic ordering of magnetic moment. From these results, we conclude that an in-plane canted antiferromagnetism is realized.

Using those weak ferromagnetic moment as a marker, we are able to investigate the local lattice distortion in thin films, which in general is very difficult to probe. Since the unit-cell volume of SrIrO$_3$ is larger than that of SrTiO$_3$[13,14], the IrO$_6$ octahedra should rotate within the plane to match the lattice constants in the superlattice. This should give rise to a DM interaction, to which the origin of the observed weak magnetic moments can be ascribed reasonably. The observed moment for $m = 1$ film ~0.02 $\mu_B$ is smaller than that of Sr$_2$IrO$_4$ 0.075 $\mu_B$[2], indicating that the smaller rotation angle. This may be reasonable since the in-plane lattice constant $a = 0.3905$ nm for the $m = 1$ sample [see Fig. 1(b)] is larger than 0.3890 nm for $Sr_2IrO_4$[2] and a smaller rotation is expected.

In bulk $Sr_2IrO_4$, the weak ferromagnetic moment is coupled antiferromagnetically along the *c*-axis and the net moment can be observed only above the meta-magnetic critical field $H_C = 0.2$ T[2,15,16]. The magnetization curve of the $m = 1$ film with clear hysteresis centered at $H = 0$, shown in Fig. 3(c), showed that the ground state is ferromagnetic rather than antiferromagnetic, implying that the interlayer alignment of weak moments are ferromagnetic, in marked contrast to bulk $Sr_2IrO_4$. To have ferromagnetic alignment of canted moments, the coupling between the adjacent $IrO_2$ layers should be either 1. the ferromagnetic alignment of Ir moments and the in-phase (ferro) rotation of $IrO_6$ octahedra, or 2. the antiferromagnetic coupling and out-of-phase rotation. We argue that the former should be the case because the interlayer coupling of Ir moments through the hybridization with inserted $SrTiO_3$ should give rise to a ferromagnetic Ir-Ir coupling regardless whether Ir-Ti coupling is ferromagnetic or antiferromagnetic. The observation of magnetic diffraction with the integer *c*-axis index [= (0.5, 0.5, 5)] indeed provides an experimental evidence for it. This leads to a magnetic and lattice ordering pattern as schematically illustrated in Fig. 4(a).

The presence of weakly ferromagnetic in-plane moments in the case of $m = 2$ implies that both the inter-bilayer and the intra-bilayer alignment of canted moments should be ferromagnetic. The ferromagnetic inter-bilayer alignment of canted moments can be described in the same way as in $m = 1$. Due to the dominant superexchange process using close to 180 degrees oxygen bonds, the bilayer magnetic coupling should be antiferromagnetic[17], which leads to the conclusion of out-of-phase rotation of $IrO_6$ octahedra between the two neighboring $IrO_2$ layers [see Fig. 4(b)]. It should be noted that, in the case of bulk bilayer $Sr_3Ir_2O_7$, the Ir moments are found to form a collinear antiferromagnetic alignment along $c$-axis[18,19], in marked contrast to the $m = 2$ superlattice. Theoretically, it was pointed out that the two states with the canted in-plane antiferromagnetism and the collinear out-of-plane antiferromagnetism are energetically almost degenerated for $Sr_3Ir_2O_7$[20], which could account for the contrasted behavior of the $m = 2$ film and $Sr_3Ir_2O_7$.

We performed first-principles calculations using density functional theory for $m = 1$ and 2 superlattices. Based on the structural analysis presented above, we assumed that the in-plane lattice constant of the heterostructure is locked to that of $SrTiO_3 \sim 0.3905$ nm, and thus the $IrO_6$ octahedra rotate about the $c$-axis in a staggered way to accommodate different bond lengths. The Ir-Ir bond length from that in orthorhombic perovskite $SrIrO_3$ (~0.394 nm)[13,14] leads a rotation angle of approximately $8^o$, which is smaller than that of bulk $Sr_2IrO_4$, $11 \sim 12^o$[15]. In Figs. 5(a) and (b) we show the band structures of $m = 1$ and 2 with small $U = 1$ eV and without magnetic ordering. SOC is already incorporated. The states near the Fermi level are Ir $t_{2g}$ states with $J_{eff} = 1/2$ character, the main player of Ir physics. The unoccupied states above ~1 eV are Ir $e_g$ and Ti $t_{2g}$ bands, implying that $SrTiO_3$ layer acts in fact insulating barrier and reducing dimensionality (See Supplementary Information for experimental verification).

For $m = 1$, $J_{eff} = 1/2$ bands around the Fermi level are narrow and simple. A half-filled character can be recognized though marginally. With increasing $m$, the interlayer interaction increases an overall band width and a gap tends to open. This gap makes the system almost band insulator but the presence of Dirac–like band of which degeneracy cannot be lifted even in the presence of SOC still remains and the system is a semimetal[5].

With introduction of a reasonably large $U = 3$ eV and magnetic order in Figs. 5(c) and (d), the evolution from a semimetal to a half-filled metal changes to the one from a weak to a strong magnetic insulator. Clearly, a relatively large gap is produced for the case of $m = 1$ but becomes marginally small for $m = 2$. Those results agree well with the experimentally observed evolution from a (semi-)metal to an insulator closely linked to a magnetism in [(SrIrO$_3$)$_m$, SrTiO$_3$] superlattice. To describe the robust insulating state with a large gap for $m = 1$, Mott like picture may be appropriate which is consistent with the insulating behavior observed even well above the magnetic transition. On the other hand, Slater-like picture may be more appropriate for $m = 2$ with a marginally small gap, where we observed much more itinerant behavior above the magnetic ordering temperature. Considering that the $m = 2$ system is a semimetal in the absence of a large $U$, it might be interesting to view the Slater-like insulator as an excitonic insulator[21] with hybridization gap produced by a magnetic ordering. We should note that the magnetic ordering pattern obtained in the calculation is a weak ferromagnet with a canted in-plane antiferromagnetic ordering within each plane both in $m = 1$ and 2, which is fully consistent with the experimental result.

In conclusion, we have materialized two dimensional slices of perovskite $SrIrO_3$ by inserting the monolayer of insulating $SrTiO_3$; $[(SrIrO_3)_m, SrTiO_3]$ ($m = 1, 2, 3, 4,$ and $\infty$) superlattices. Utilizing a delicate interplay of very strong SOC, electron correlations and dimensionality, a semimetal-magnetic insulator transition and spin-orbital Mott state was realized. Likely due to relatively weak correlation, the electronic state can be well predicted by first-principles calculation on quantitative level, which implies that we may tailor a variety of exotic states in the novel family of complex $5d$ transition metal oxides. The apparent target should be the various topological states proposed recently[5–8,11] and this work should be an important step towards the goal.

**Methods**

**Sample fabrication.** The superlattice samples were fabricated on $SrTiO_3$(001) substrates by a pulsed laser deposition technique using KrF excimer laser pulses focused on $SrIrO_3$ ceramic target and $SrTiO_3$ single crystal target. The laser fluence and the repetition were 1-2 J/cm$^2$ and 2 Hz, respectively. The typical substrate temperature was 620°C and the oxygen pressure was 25 Pa. The number of superlattice units, $k$, is designed so as to make the total thickness about 80 perovskite unit cells.

**Measurements.** Resonant magnetic x-ray diffraction experiments were performed at beamline 3A at Photon Factory, KEK. Photon polarization of the σ-π' channel was selectively measured by using a Mo(400) analyzer crystal.

**Theoretical calculations.** The full potential linear augmented plane-wave method as implemented in the elk code[22] was used to perform calculations to incorporate SOC and Hubbard $U$. We employed the Ceperley-Alder form[23] for the exchange-correlation functional within the local density approximation (LDA) as parametrized by Perdew and Zunger[24]. SOC was included in the second-variational step, and the around mean-field scheme[25] for the double-counting corrections was used. All calculations used a grid size of $4 \times 4 \times 4$ for Brillouin zone integration, and the number of empty states for the second variational step was 12. Muffin tin radii (in atomic units) of 2.0 for Ir and Ti, 1.6 for O, and 2.2 for Sr were used.

**Acknowledgements**

This work was supported by a Grant-in-Aid for Scientific Research (No. 19104008, 22340108, 25103724) from MEXT, Japan. We would like to thank Y. Yamazaki, H. Nakao, and Y. Murakami for helpful experimental support at Photon Factory, KEK. Hard x-ray diffraction measurements were performed under the approval of the Photon Factory Program Advisory Committee (Proposals No. 2011G062) at the Institute of Material Structure Science, KEK. V.V.S. and H.Y.K are supported by NSERC of Canada. Computations were performed on GPC supercomputer at the SciNet HPC Consortium. SciNet is funded by: the Canada Foundation for Innovation under the auspices of Compute Canada; the Government of Ontario; Ontario Research Fund - Research Excellence; and the University of Toronto.

**Author contributions**

J.M. designed the experiments. J.M. and K.I. fabricated samples and collected the basic data. S.Y., H.W. and K.I performed the experiment using synchrotron radiation. V.V.S. and H.Y.K. performed theoretical calculations. J.M. wrote the manuscript with input from H.Y.K. and H.T. H.T. planned and supervised the project. All authors discussed the results.

**Additional information**

The authors declare no competing financial interests.

**Figure legends**

**Figure 1. Structural characterization of atomically controlled artificial superlattice consisting of SrIrO₃ and SrTiO₃.** **a,** X-ray diffraction scans (θ-2θ) of the superlattices [(SrIrO$_3$)$_m$, SrTiO$_3$]$_k$/SrTiO$_3$(001). Downarrows indicate the superlattice peaks, while open circles represent the fundamental peaks. Open triangles depict peaks from the substrate. **b,** Reciprocal-space mapping of the (114) x-ray diffraction peaks for the superlattice with $m = 1$ grown on SrTiO$_3$(001) substrate. **c,** Atomically resolved HAADF-STEM image of the superlattice with $m = 1$ along SrTiO$_3$[100] direction. The schematic of crystal structure is superposed on the image to show the brightest spot (Ir) is atomically aligned in (001) plane.

**Figure 2. Physical properties of the superlattice samples with controlling dimensionality.** **a,** Schematics of the superlattices with $m = 1$, 2, and ∞, demonstrating the controlled dimensionality. **b,** Resistivity (ρ), **c,** Hall coefficient ($R_H$), and **d,** in-plane magnetization ($M$) are plotted as a function of temperature $T$. Arrows in each panel indicate temperatures of anomaly in the resistivity for $m = 1$ and 2. The anomaly is defined as a peak temperature of a second derivative of log(ρ).

**Figure 3. Magnetization of the superlattice with $m = 1$. a,** Temperature ($T$) dependence of the magnetization ($M$) measured at 0.1 T. Magnetization was measured for magnetic field ($H$)

perpendicular to the IrO$_2$ plane and parallel to the IrO$_2$ plane. The upturn below 50 K, observed for both magnetic field directions, is derived from magnetic impurities in the substrates. **b,** Temperature dependence of the magnetic x-ray diffraction intensity for (0.5, 0.5 ,5) peak. The unit cell dimensions are $a \times a \times 2a$ ($a \sim 0.39$ nm). **c,** Field dependence of the in-plane magnetization measured at 100 K.

**Figure 4. Magnetic structures.** The superlattices with **a,** $m = 1$ and **b,** $m = 2$ and **c,** Sr$_2$IrO$_4$[2]. Phase relation of the IrO$_6$ rotation between adjacent IrO$_2$ layers is determined so as to account for the observed magnetic moments. While we have no experimental information on TiO$_6$ rotation, we assumed it to be out of phase with the neighboring IrO$_6$ rotation since it is reasonable to avoid Coulomb repulsion between oxygen atoms.

**Figure 5. Calculated electronic structures of the $m = 1$ and 2 superlattices.** Band structures for **a,** $m = 1$ and **b,** $m = 2$ with $U = 1$ eV. The red and blue colors represent Ir and Ti $d$-orbitals, respectively. The grey lines dominant below -1.5 eV indicates oxygen $p$-orbitals. The dashed line denotes the Fermi level, and the bands near the Fermi level are mostly $J_{\text{eff}} = 1/2$ character except near $\Gamma$ and Z points. Band structures in the presence of a magnetic order for **c,** $m = 1$ and **d,** $m = 2$ with $U = 3$ eV. Both manifest a charge gap shaded in orange.

**Figures**

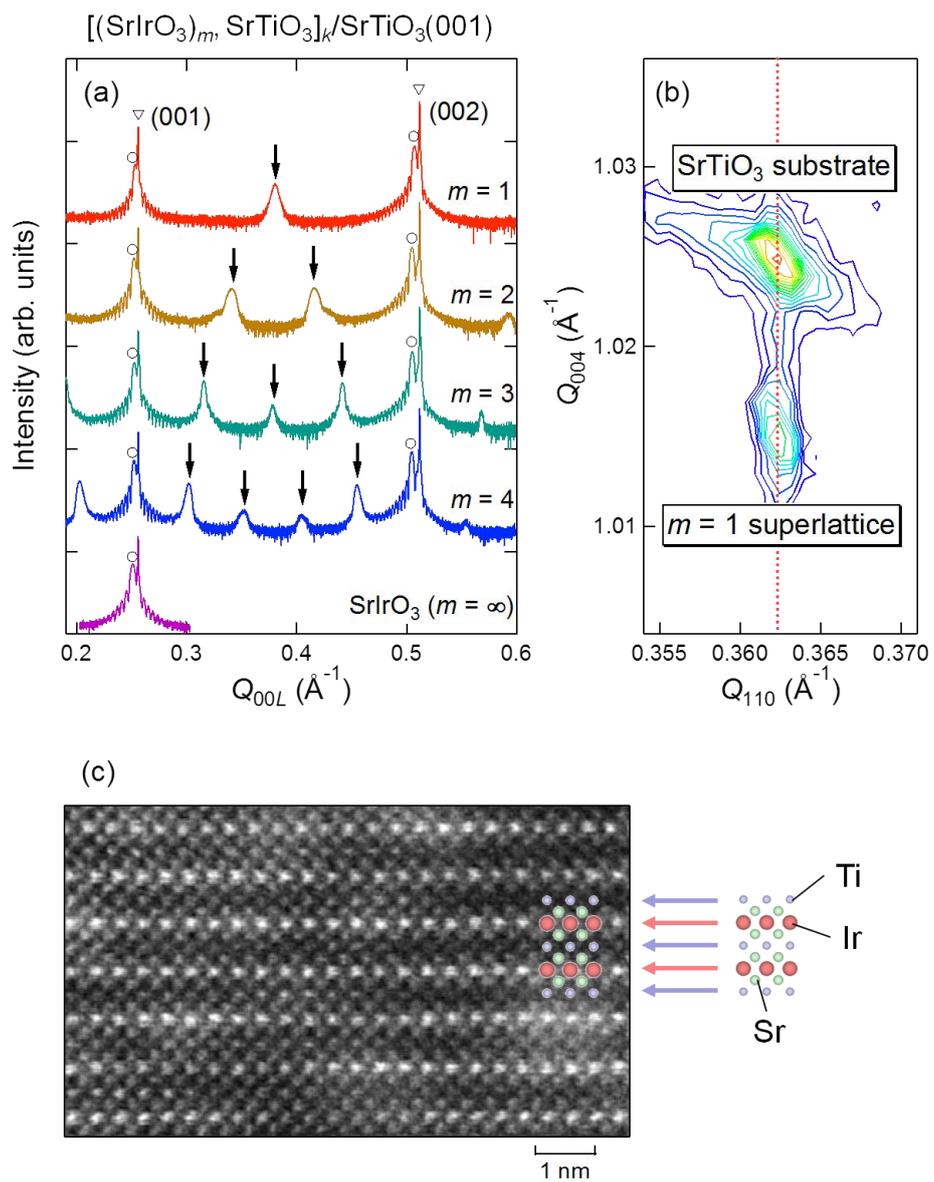

Fig. 1     J. Matsuno *et al.*

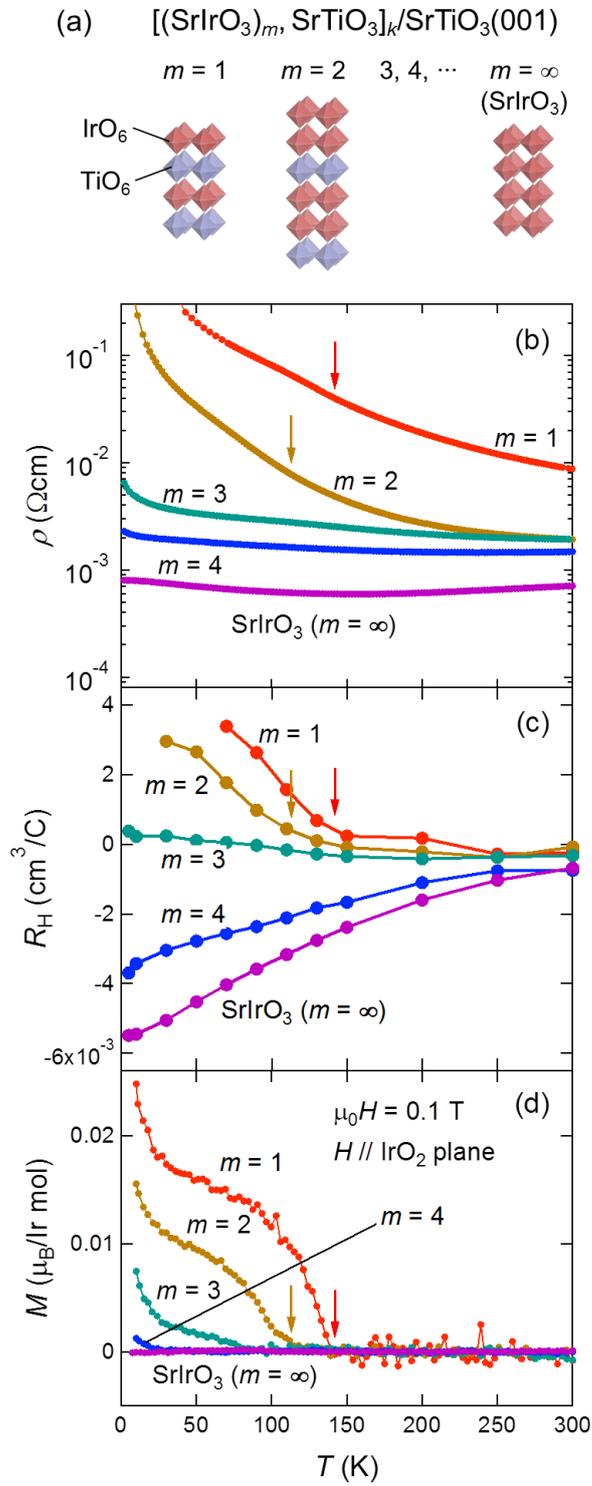

Fig. 2　　J. Matsuno *et al.*

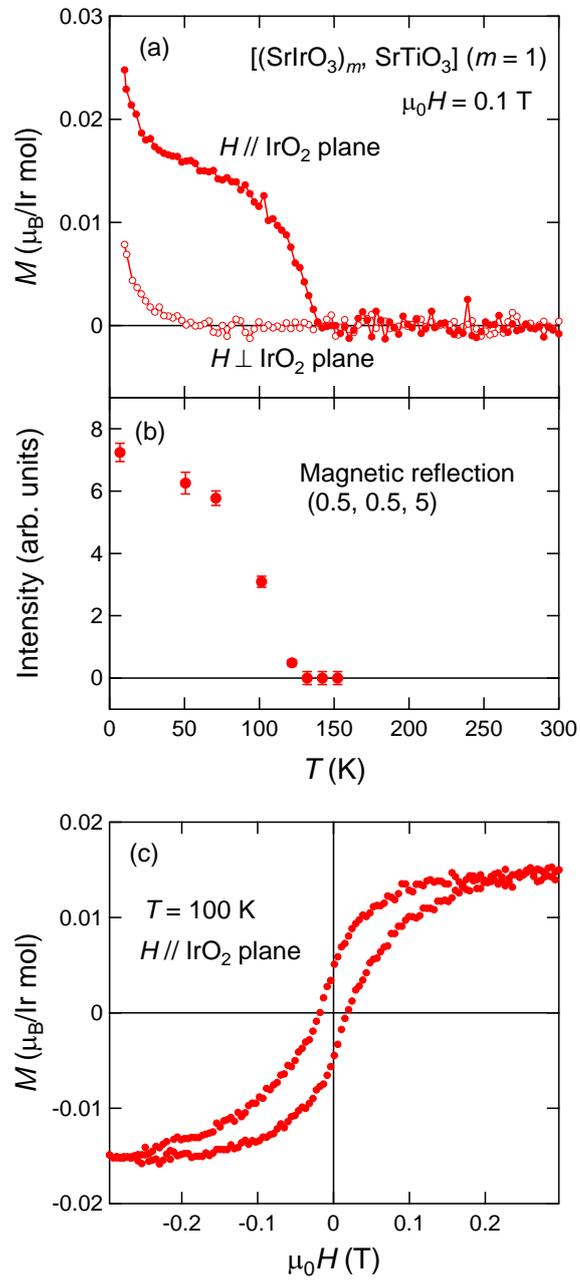

Fig. 3   J. Matsuno *et al.*

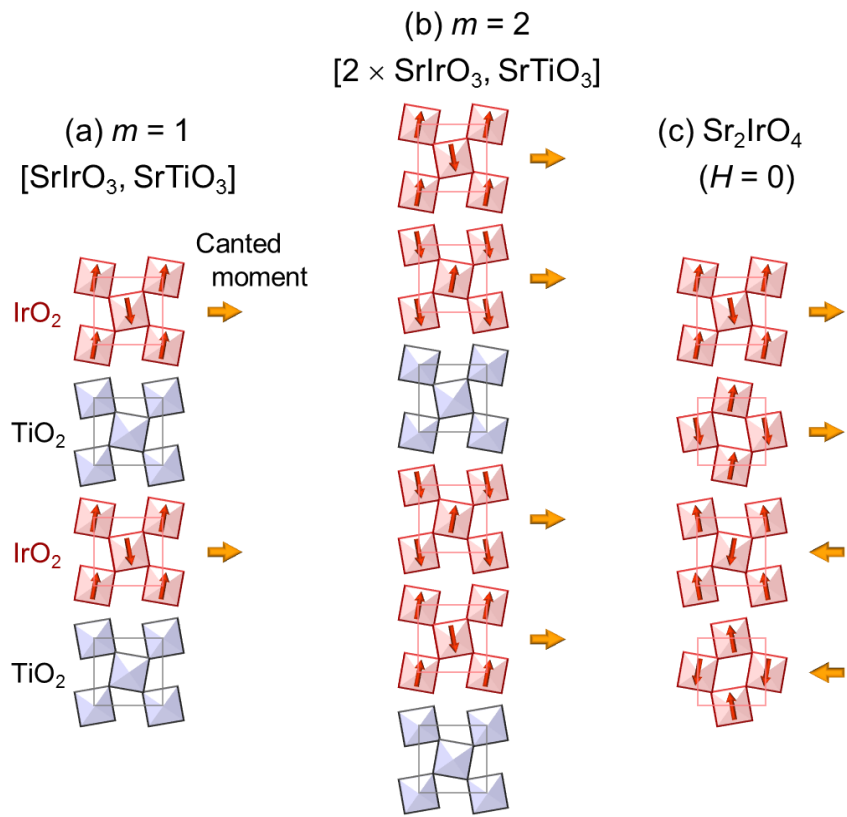

Fig. 4  J. Matsuno *et al.*

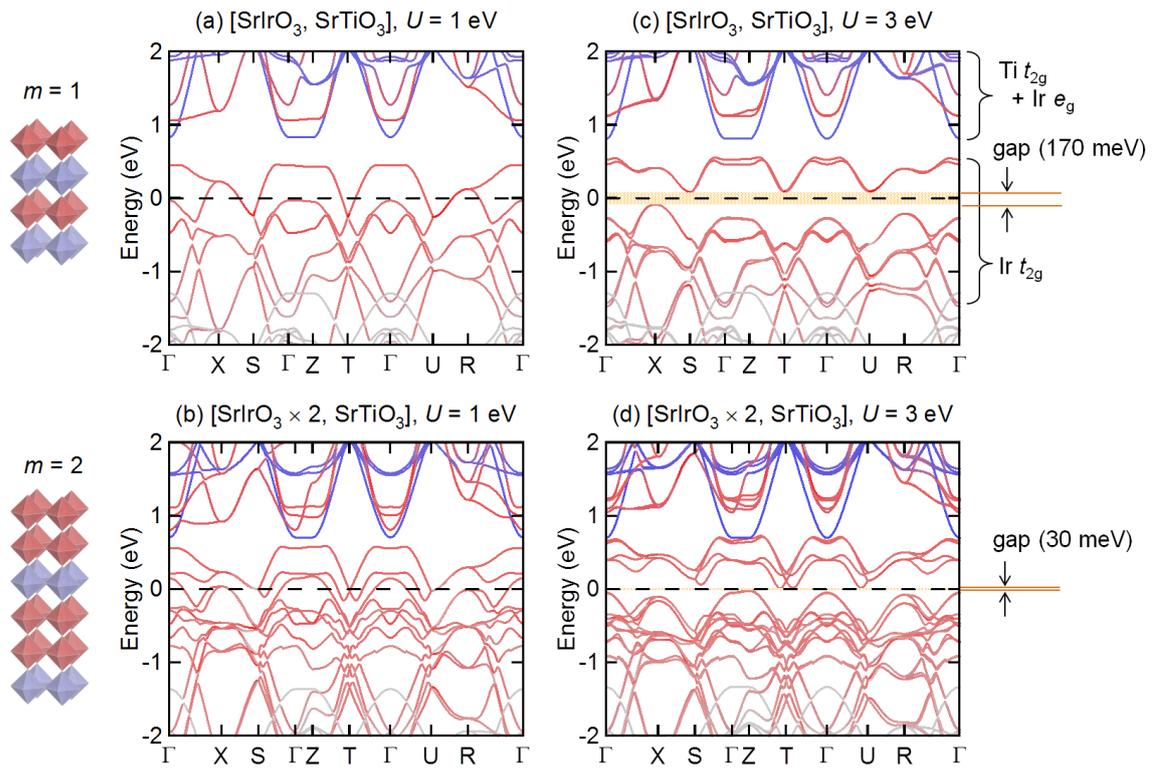

Fig. 5    J. Matsuno *et al.*

# SUPPLEMENTARY INFORMATION

# Engineering spin-orbital magnetic insulator by tailoring superlattices

**Electronic structure of $SrIrO_3$/$SrTiO_3$ superlattices experimentally confirmed by photoemission and x-ray absorption spectroscopies**

In order to experimentally confirm the electronic structures of the superlattices, we performed hard x-ray photoemission spectroscopy (HXPES) and O 1$s$ x-ray absorption spectroscopy (XAS), corresponding to occupied and unoccupied density of states, respectively. The results for $m = 1$, 4, and $\infty$ are indicated in Fig. S1. In the valence-band photoemission (left), the Ir 5$d$ band is located near the Fermi level ($E_F$), from 0 to -2 eV, and the O 2$p$ band is located on the lower-energy side from -4 to -9eV. The line shape Ir 5$d$ band of $m = 1$ superlattice is different from the others in that the intensity at $E_F$ is weaker as well as the structure is shifted to a lower energy; the electronic structure of the $m = 1$ is possibly dominated by Coulomb repulsion, which can drive the density of states out of $E_F$. In O 1$s$ XAS, the Ir 5$d$ and Ti 3$d$ bands are located at about 530 eV and 532 eV, respectively. The intensities of Ir 5$d$ bands increase as $m$ increases from 1 to $\infty$. From these results, the electronic states in the vicinity of $E_F$ are solely occupied by Ir 5$d$ bands whereas Ti 3$d$ bands are well above $E_F$ and hence unoccupied. Namely, the valence of Ir in the $SrIrO_3$ layer is 4+, as in $Sr_2Ir^{4+}O_4$ ($5d^5$), while the $SrTi^{4+}O_3$ ($3d^0$) layer can be regarded as a blocking layer. Those findings strongly support the calculated orbital character shown in Fig. 5.

**Methods**

We performed HXPES at the beamline BL-47XU of SPring-8[26], and O 1$s$ XAS at the beamline BL-2C of Photon Factory, high Energy Accelerators Research Organization (KEK). Photon energy and energy resolution are 7938 eV and 200 meV in HXPES, and 520 - 560 eV and 100 meV in XAS, respectively. All the measurements were performed at room temperature. In HXPES, the position of the Fermi level ($E_F$) was determined by measuring the spectra of the gold, which is in electrical contact with the superlattices. In O 1$s$ XAS, the experimental geometry is also shown in Fig. S1. The incident angle of the incoming x-ray with horizontal linear polarization was 60° from the sample surface normal.

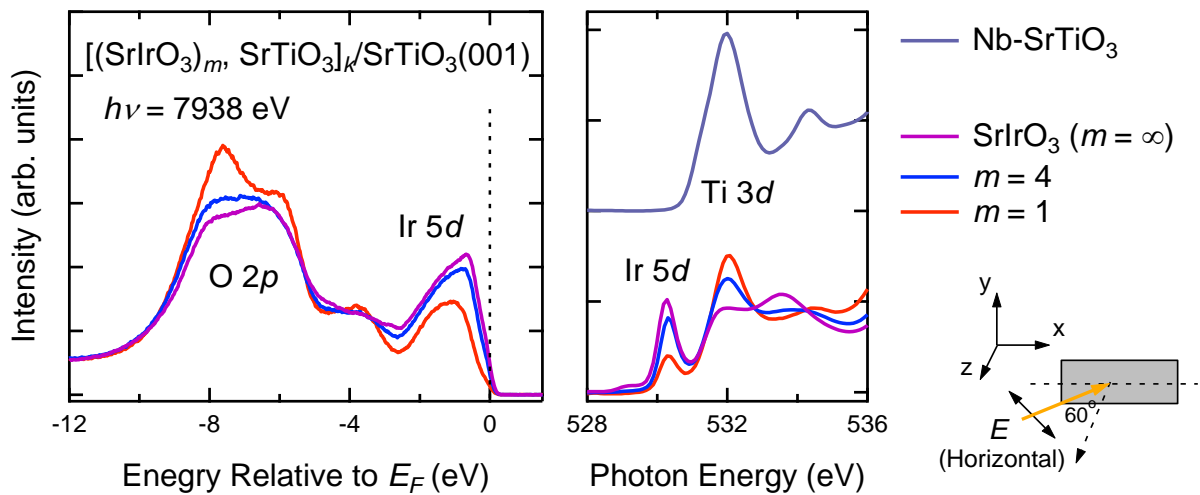

**Figure. S1 Electronic structure of the superlattice samples with $m = 1$, 4, and $\infty$.**

Valence-band photoemission (left) and the O 1$s$ XAS (right) spectra of [(SrIrO$_3$)$_m$, SrTiO$_3$] superlattices. For the XAS, the spectrum of Nb-doped SrTiO$_3$ is also plotted as a reference. The experimental geometry of O 1$s$ XAS is also shown.

**Acknowledgements**

We would like to thank T. Sugiyama, E. Ikenaga, and H. Kumigashira for helpful support for the experiments. The HXPES measurements at SPring-8 were performed under the approval of the Japan Synchrotron Radiation Research Institute (Proposals No. 2011A1624, No. 2011B1710, No. 2012A1624 and No. 2012B1757). The O 1$s$ XAS measurements were performed under the approval of the Photon Factory Program Advisory Committee (Proposals No. 2011G061 and No. 2011S2-003) at the Institute of Material Structure Science, KEK.